\documentclass[conference]{IEEEtran}
\usepackage{wrapfig}
\usepackage{amsfonts}
\usepackage{amsmath}
\usepackage[utf8]{inputenc}
\usepackage[T1]{fontenc}
\usepackage{graphicx}
\usepackage[english]{babel}
\usepackage[algoruled]{algorithm2e}

\renewcommand{\theequation}{\thesection.arabic{equation}}

\renewcommand{\thefigure}{\thesection.\arabic{figure}}

\renewcommand{\vec}[1]{\mathbf{#1}}
\renewcommand{\theequation}{\thesubsection.\arabic{equation}}
\DeclareGraphicsExtensions{.pdf,.png,.jpg, .gif}

\usepackage{amsthm}

\usepackage[english]{babel}
\usepackage{mathtools}

\DeclareFontFamily{U}{wncy}{}
\DeclareFontShape{U}{wncy}{m}{n}{<->wncyr10}{}
\DeclareSymbolFont{mcy}{U}{wncy}{m}{n}
\DeclareMathSymbol{\Sh}{\mathord}{mcy}{"58} 
\DeclareMathOperator*{\argmin}{arg\,min}

\newcounter{eqn}

\makeatother
\newcommand{\vectornorm}[1]{\left|\left|#1\right|\right|}

\newtheorem{thm}{Theorem}
\newtheorem{defn}{Definition}
  \newtheorem{theorem}{Theorem}[section]

  \newtheorem{lemma}[theorem]{Lemma}

\newcommand{\re}{{\mathbb{R}}}

\newcommand{\pr}{{\mathbb{P}}}

\title{Distributed Wideband Spectrum Sensing}

\author{Thomas Kealy$^{1}$, Oliver Johnson$^{2}$ and Robert Piechocki$^{3}$
\thanks{This work was supported by the Engineering and Physical Sciences Research Council [grant number {\tt EP/I028153/1}]; Ofcom; and the University of Bristol. The authors would particularly like to thank Gary Clemo of Ofcom for useful discussions.}%
\thanks{ $^{1}$Thomas Kealy is with CDT in Communications, MVB, School of Engineeering University of Bristol, UK
        {\tt\small tk12098@bristol.ac.uk}}%
\thanks{$^{2}$Oliver Johnson is with the Department of Mathematics, University Walk, Bristol, University of Bristol, UK.
        {\tt\small O.Johnson@bristol.ac.uk}}%
\thanks{$^{3}$Robert Piechocki is with the CSN Group, MVB, School of Engineering, University of Bristol, UK.
        {\tt\small r.j.piechocki@bristol.ac.uk}}%
}

\IEEEoverridecommandlockouts

\begin{document}

\maketitle

\begin{abstract}
\noindent We consider the problem of reconstructing wideband frequency spectra from distributed, compressive measurements. The measurements are made by a network of nodes, each independently mixing the ambient spectra with low frequency, random signals. The reconstruction takes place via local transmissions between nodes, each performing simple statistical operations such as ridge regression and shrinkage.
\end{abstract}

\section{Introduction}

There is an almost ubiquitous growing demand for mobile and wireless data, with consumers demanding faster speeds and better quality connections in more places. Consequently 4G is now being rolled out in the UK and US and with 5G being planned for 2020 and beyond \cite{Dahlman2014}.  

However, there is constrained amount of frequencies over which to transmit this information; and demand for frequencies that provide sufficient bandwidth, good range and in-building penetration is high.

Not all spectrum is used in all places and at all times, and judicious spectrum management, by developing approaches to use white spaces where they occur, would likely be beneficial.

Broadly, access to spectrum is managed in two, complementary ways, namely through licensed and licence exempt access. Licensing authorises a particular user (or users) to access a specific frequency band. Licence exemption allows any user to access a band provided they meet certain technical requirements intended to limit the impact of interference on other spectrum users.

A licence exempt approach might be particularly suitable for managing access to white spaces. Devices seeking to access white spaces need a robust mechanism for learning of the frequencies that can be used at a particular time and location. One approach is to refer to a database, which maps the location of white spaces based on knowledge of existing spectrum users. An alternative approach is for devices to detect white spaces by monitoring spectrum use. 

The advantages of spectrum monitoring \cite{akan2009cognitive} over persisting a database of space-frequency data are the ability of networks to make use of low-cost low-power devices, only capable of making local (as opposed to national) communications, keeping the cost of the network low and  opportunistic channel usage for bursty traffic, reducing channel collisions in dense networks.

The realisation of any Cognitive Radio standard (such as IEEE 802.22 \cite{stevenson2009ieee}), requires the co-existence of primary (TV users) and secondary (everybody else who wants to use TVWS spectrum) users of the frequency spectrum to ensure proper interference mitigation and appropriate network behaviour. We note, that whereas TVWS bands are an initial step towards dynamic spectrum access, the principles and approaches we describe are applicable to other frequency bands - in particular it makes ultra-wideband spectrum sensing possible.

The challenges of this technology are that Cognitive Radios (CRs) must sense whether spectrum is available, and must be able to detect very weak primary user signals. Furthermore they must sense over a wide bandwidth (due to the amount of TVWS spectrum proposed), which challenges traditional Nyquist sampling techniques, because the sampling rates required are not technically feasible with current RF or Analogue-to-Digital conversion technology.

Due to the inherent sparsity of spectral utilisation, Compressive Sensing (CS) \cite{Candes2006} is an appropriate formalism within which to tackle this problem. CS has recently emerged as a new sampling paradigm allowing images to be taken from a single pixel camera for example. Applying this to wireless communication, we are able to reconstruct sparse signals at sampling rates below what would be required by Nyquist theory, for example the works \cite{mishali2010theory}, and \cite{tropp2010beyond} detail how this sampling can be achieved. 

However, even with CS, spectrum sensing from a single machine will be costly as the proposed TVWS band will be over a large frequency range (for instance in the UK the proposed TVWS band is from 470 MHz to 790 MHz, requiring traditional sampling rates of \textasciitilde 1600 MHz). CS at a single sensor would still require high sampling rates. In this paper we propose a distributed model, which allows a sensing budget at each node far below what is required by centralised CS. The main contribution of this paper is that the model can be solved in a fully distributed manner - we do not require a central fusion centre as in \cite{Zhang2011b}. Moreover, we are able to show that the set of updates at each nodes takes closed form.

The structure of the paper is as follows: in section \ref{sec:sensingmodel} we introduce the sensing model, in section \ref{sec:opt-on-graphs} we describe the distributed reconstruction algorithm \cite{mota2013d}, and finally in section \ref{sec:results} we show some results of the reconstruction quality of this model. 

\section{Model}\label{sec:sensingmodel}

We consider a radio environment with a single primary user (PU) and a network of \(J\) nodes collaboratively trying to sense and reconstruct the PU signal, either in a fully distributed manner (by local communication), or by transmitting measurements to a fusion centre which then solves the linear system. 

We try to sense and reconstruct a wideband signal, divided into \(L\) channels. We have a (connected) network of \(J\) (= 50) nodes placed uniformly at random within the square \(  \left[0,1\right]\times \left[0,1\right] \). This is the same model, as in \cite{Zhang2011b}. The calculations which follow are taken from \cite{Zhang2011b} as well.

The nodes individually take measurements (as in \cite{mishali2010theory}) by mixing the incoming analogue signal \(x\left(t\right)\) with a mixing function \(p_i\left(t\right)\) aliasing the spectrum. \(x\left(t\right)\) is assumed to be bandlimited and composed of up to \(k\) uncorrelated transmissions over the \(L\) possible narrowband channels - i.e. the signal is \(k\)-sparse. 

The mixing functions - which are independent for each node - are required to be periodic, with period \(T_p\). Since \(p_i\) is periodic it has Fourier expansion:

\begin{equation}
p_i\left(t\right) = \sum_{l=-\infty}^{\infty} c_{il} \exp\left({jlt\frac{2\pi}{T_p}}\right)
\end{equation}

The \(c_{il}\) are the Fourier coefficients of the expansion and are defined in the standard manner. The result of the mixing procedure in channel \(i\) is therefore the product \(xp_i\), with Fourier transform (we denote the Fourier Transform of \(x\) by \(X\left( \dot{.} \right)\)):

\begin{align}
X_{i}\left(f\right) &=& \int_{-\infty}^{\infty} x\left(t\right) p_i\left(t\right) dt \nonumber
\\ &=& \sum_{l=-\infty}^{\infty} c_{il} X\left(f-lf_p\right)
\end{align}

(We insert the Fourier series for \(p_i\), then exchange the sum and integral). The output of this mixing process then, is a linear combination of shifted copies of \(X\left(f\right)\), with at most \(\lceil f_NYQ/f_p\rceil\) terms since \(X\left(f\right)\) is zero outside its support (we have assumed this Nyquist frequency exists, even though we never sample at that rate).

This process is repeated in parallel at each node so that each band in \(x\) appears in baseband.

Once the mixing process has been completed the signal in each channel is low-pass filtered and sampled at a rate \(f_s \geq f_p\). In the frequency domain this is a ideal rectangle function, so the output of a single channel is:

\begin{equation}
Y_i\left(e^{j 2 \pi f T_s }\right) = \sum_{l = -L_0}^{+L_0} c_{il} X\left(f-lf_p\right)
\end{equation}

since frequencies outside of \([-f_s/2, f_s/2]\) will filtered out. \(L_0\) is the smallest integer number of non-zero contributions in \(X\left(f\right)\) over \([-f_s/2, f_s/2]\) - at most \(\lceil f_NYQ/f_p\rceil\) if we choose \(f_s = f_p\). These relations can be written in matrix form as:

\begin{equation}
\textbf{y} = \textbf{A}\textbf{x} + \vec{w}
\label{system}
\end{equation}

where \(\textbf{y}\) contains the output of the measurement process, and \(\textbf{A}\) is a product matrix of the mixing functions, their Fourier coefficients, a partial Fourier Matrix, and a matrix of channel coefficients. \(\textbf{x}\) is the vector of unknown samples of \(x\left(t\right)\). 

i.e. \(\textbf{A}\) can be written: 

\begin{equation}
\textbf{A}^{m\times L} = \textbf{S}^{m\times L} \textbf{F}^{L\times L} \textbf{D}^{L \times L} \textbf{H}^{L \times L}
\end{equation}

The measurements \(\textbf{y}\) are transmitted to a Fusion Centre via a control channel. The system  \ref{system} can then be solved (in the sense of finding the sparse vector \(\vec{x}\) by convex optimisation via minimising the objective function:

\begin{equation}
\frac{1}{2}\|\textbf{Ax}-\textbf{y}\|_2^2 + \lambda \|\textbf{x}\|_1
\end{equation}

where \(\lambda\) is a parameter chosen to promote sparsity. Larger \(\lambda\) means sparser \(\vec{x}\).

\section{ADMM }\label{sec:admm}
The alternating direction method of multipliers \cite{Boyd2010a}, (ADMM), algorithm solves problems of the form

\begin{align}
\argmin_{x} f\left( x \right) + g\left(z\right) \nonumber
\\
\text{s.t } Ux +Vz = c
\label{admm}
\end{align}

where \(f\) and \(g\) are assumed to be convex function with range in \(\re\), \(U \in \re^{p \times n}\) and \(V\in \re^{p \times m}\) are matrices (not assumed to have full rank), and \(c \in \re^p\).

ADMM consists of iteratively minimising the augmented Lagrangian 

\begin{align*}
L_p\left(x, z, \eta\right) = f\left( x\right) +& g\left(z\right)+\eta^T\left(Ux+Vz-c\right) + \\ \frac{\rho}{2}\|Ux+Vz-c\|_2^2
\label{admm_form}
\end{align*}

(\(\eta\) is a Lagrange multiplier), and \(\rho\) is a parameter we can choose to make \(g(z)\) smooth \cite{nesterov2005smooth}, with the following iterations:

\begin{align}
x^{k+1} &:= \argmin_{x} L_\rho\left(x,z^k,\eta^k\right)\\
z^{k+1} &:= \argmin_{z} L_\rho\left(x^{k+1},z,\eta^k\right)\\
\eta^{k+1} &:= \eta^{k} + \rho \left(Ux^{k+1} + Vz^{k+1} - c\right)
\label{admm_algo}
\end{align}

The alternating minimisation works because of the decomposability of the objective function: the \(x\) minimisation step is independent of the \(z\) minimisation step and vice versa.  

We illustrate an example, relevant to the type of problems encountered in signal processing.

\subsection{Example: ADMM for Centralised LASSO}
ADMM can be formulated as an iterative MAP estimation procedure for the problem (which is referred to as LASSO see \cite{tibshirani1996regression}):

\begin{eqnarray}
\frac{1}{2}\|Ax-b\|_2^2 + \lambda\|x\|_1
\end{eqnarray}

This can be cast in constrained form as:

\begin{eqnarray}
\frac{1}{2}\|Ax-b\|_2^2 + \lambda\|z\|_1 \\
\text{s.t } z = x
\end{eqnarray}

i.e this is of the form \eqref{admm} with \( f\left(x\right) =\|Ax-y\|_2^2\), \(g\left(z\right) = \lambda\|z\|_1\), \(U=I\), \(V=-I\), and \(c=0\).

The associated Lagrangian is:

\begin{equation}
L_\rho = \frac{1}{2}\|Ax-b\|_2^2 + \lambda\|z\|_1 + \eta\left(x-z\right) + \frac{\rho}{2}\|x-z\|^2
\label{eq:lasso-lagrangian}
\end{equation}

Now, given a set of noisy measurements (say of radio spectra) \(\vec{y}\), and a sensing matrix \(\vec{A}\) we can use ADMM to find the (sparse) radio spectra.

The ADMM iterations for LASSO, which can be found by alternately differentiating \eqref{eq:lasso-lagrangian} with respect to \(x\),\(z\) and \(\eta\), are (in closed form):

\begin{align}
x^{k+1} &:= \left(A^TA + \rho I\right)^{-1}\left(A^Tb +\rho\left(z^k - y^k\right)\right)\\
z^{k+1} &:= S_{\lambda/\rho}\left(x^{k+1} + \eta^k/\rho\right)
 \\
\eta^{k+1} &:= \eta^{k} + \rho \left(x^{k+1}-z^{k+1}\right)
\label{admm_algo_lasso}
\end{align}

where \(S_{\lambda/\rho}\left(\circ\right)\) is the soft thresholding operator: \(S_\gamma\left(x\right)_i = \mathrm{sign}(x_i)\left(|x_i| - \gamma\right)^+\).

This algorithm has a nice statistical interpretation: it iteratively performs ridge regression, followed by shrinkage towards zero. This is the MAP estimate for \(x\) under a Laplace prior.

The soft-thresholding operator can be derived by considering the MAP estimate of the following model:

\begin{equation}
y = x + w
\end{equation}

where \(x\) is some (sparse) signal, and \(w\) is additive white Gaussian noise. We seek

\begin{equation}
\hat{x} = \arg\max_x \pr_{x|y}{\left(x|y\right)}
\end{equation}

This can be recast in the following form by using Bayes rule, noting that the denominator is independent of \(x\) and taking logarithms:

\begin{equation}
\hat{x} = \arg\max_x \left[\log{\pr_{w}{\left(y-x\right)}}+\log{\pr{\left(x\right)}}\right]
\label{hatx}
\end{equation}

The term \(\pr_{n}{\left(y-x\right)}\) arises because we are considering \(x+w\) with \(w\) zero mean Gaussian, with variance \(\sigma_n^2\). So, the conditional distribution of \(y\) (given \(x\)) will be a Gaussian centred at \(x\).

We will take \(\pr{\left(x\right)}\) to be a Laplacian distribution:

\begin{equation}
\pr{\left(x\right)} = \frac{1}{\sqrt{2}\sigma}\exp{-\frac{\sqrt{2}}{\sigma}|x|}
\end{equation}

Note that \( f\left(x\right) = \log{\pr_x{ \left( x \right)}} ~ -\frac{\sqrt{2}}{\sigma} |x| \), and so by differentiating \( f'\left(x\right) = -\frac{\sqrt{2}}{\sigma} \mathrm{sign}\left(x\right) \)

Taking the maximum of \ref{hatx} we obtain:

\begin{equation}
\frac{y-\hat{x}}{\sigma^2_n}-\frac{\sqrt{2}}{\sigma}sign(x) = 0
\end{equation}

Which leads the soft thresholding operation defined earlier, with \(\gamma = \frac{\sqrt{2}\sigma^2_n}{\sigma}\) as (via rearrangement):

$$
y =  \hat{x} + \frac{\sqrt{2}\sigma^2_n}{\sigma}\mathrm{sign}\left(x\right)
$$

or

$$
\hat{x}\left(y\right) = \mathrm{sign}(y)\left(y - \frac{\sqrt{2}\sigma^2_n}{\sigma}\right)_+
$$

i.e \(S_\gamma(y)\).

\section{Constrained Optimisation on Graphs}\label{sec:opt-on-graphs}
We model the network as an undirected graph \(G = \left(V,E\right)\), where \(V = \{1 \ldots J\}\) is the set of vertices, and \(E = V \times V\) is the set of edges. An edge between nodes \(i\) and \(j\) implies that the two nodes can communicate. The set of nodes that node \(i\) can communicate with is written \(\mathcal{N}_i\) and the degree of node \(i\) is \(D_i = |\mathcal{N}_i|\). 

Individually nodes make the following measurements:

\begin{equation}
\vec{y}_p = \vec{A}_p\vec{x} + \vec{n}_p
\end{equation}

where \(\vec{A}_p\) is the \(p^{th} \) row of the sensing matrix from \eqref{system}, and the system \eqref{system} is formed by concatenating the individual nodes' measurements together.

We assume that a proper colouring of the graph is available: that is, each node is assigned a number from a set \(C = \{1 \ldots c \} \), and no node shares a colour with any neighbour.

To find the \(\vec{x}\) we are seeking, to each node we give a copy of \(\vec{x}, \vec{x}_p\) and we constrain the copies to be indentical across all edges in the network. We can write the combined optimisation variable as \(\bar{x}\), which collects together \(C\) copies of a \(n\times 1\) vector \(\vec{x}\):

\begin{defn}
We define vectors \(x_c\), where \(c = 1,\ldots , C\) and write the vector of length \(nJ\):
\begin{equation}
\bar{x} = \sum_{c=1}^C w_c \otimes x_c = \left[x_{c(1)}^T, \ldots	, x_{c(J)}^T\right]^T
\label{barxc}
\end{equation}
where \(w_{c(i)} = \mathbb{I}(c(i) = c)\), \(\mathbb{I}\) is the indicator function, and we have written \(c(i)\) for the colour of the \(i\)th node.
\end{defn}

The problem then is to solve:

\begin{align}
\argmin_{\bar{x}} \sum_{c=1}^C \sum_{j \in c} \|A_jx_j - y_j\|_2^2 + \frac{\lambda}{J}\|x\|_1 \nonumber \\ 
\text{ and } x_i = x_j \text{ if } \{i,j\} \in E \nonumber \\
\text{ and } x_i = z_i \text{ } \forall i \in \{1, \ldots, C\}
\label{constrainedbp}
\end{align}

\begin{figure}[h]
\centering
\includegraphics[height = 5 cm]{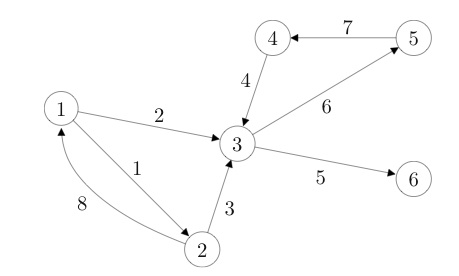}
\caption{An example of a network}
\label{efig:ex-network}
\end{figure}

\begin{figure}[h]
\centering
\includegraphics[height = 3 cm, width = 7cm]{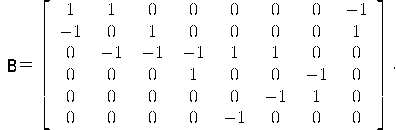}
\caption{The incidence matrix associated with Figure \eqref{efig:ex-network}}
\label{fig:incidence-matrix}
\end{figure}

These constraints can be written more compactly by introducing the node-arc incidence matrix B: a \(V\) by \(E\) matrix where each column is associated with an edge \(\left(i,j\right) \in E\) and has \(1\) and \(-1\) in the \(ith\) and \(jth\) entry respectively. Figures \eqref{efig:ex-network} and \eqref{fig:incidence-matrix} show examples of a network and it's associated incidence matrix.

\begin{defn}
\begin{align*}
u &:= \left(B^T \otimes I_n\right)\bar{x} \\
& = \left(B^T \otimes I_n\right)\sum_{c=1}^C w_c \otimes x_c \\
& = \sum	_{c=1}^C B_c^T\otimes x_c
\end{align*}
where we have used the definition \eqref{barxc} in the second line, and the property of Kronecker products \((A\otimes C)(B \otimes D) = (AB \otimes CD)\) between the second and third lines, and we write \(B_c = w_c^TB\).
\end{defn}

The constraint \(x_i = x_j \text{ if } \{i,j\} \in E \) can now be written 

\begin{equation}
\sum_{c=1}^C\left(B_c^T \otimes I_n\right)\bar{x}_c = 0
\label{compact-constraints}
\end{equation}

note that \(\left(B^T\otimes I_n \right) \in \re^{nE \times nJ}\). Together \eqref{barxc} and \eqref{compact-constraints}, suggests that the problem \eqref{constrainedbp} can be re-written as:

\begin{align}
\argmin_{\bar{x}} \sum_{c=1}^C \sum_{j \in C_c} \|A_jx_j - y_j\|_2^2 + \beta\|z_j\|_1
\nonumber \\
\text{ s.t. } \sum_{c=1}^C\left(B_c^T \otimes I_n\right)\bar{x}_c = 0 \nonumber \\
\text{ and } \bar{x}_c - \bar{z}_c = 0
\label{constrainedbp1}
\end{align}

where \(\beta = \frac{\lambda}{J}\).

The Augmented Lagrangian for the problem \eqref{constrainedbp1} can be written down as:

\begin{align}
L_\rho = \sum_{c=1}^C ( \sum_{j \in c} & \left(\|A_jx_j - y_j\|_2^2 + \beta\|z_j\|_1 \right) + \nonumber \\ &\eta^T\left(B_c^T \otimes I_n\right)\bar{x}_c +  \frac{\rho}{2}\vectornorm{\bar{x}_c-\bar{z}_c}_2^2  + \nonumber \\  &\theta^T\left(\bar{x}_c - \bar{z}_c\right)  + \frac{\rho}{2}\vectornorm{\sum_{c=1}^C\left(B_c^T \otimes I_n\right)\bar{x}_c}_2^2
\label{aug-lagrange}
\end{align}

The term (\(\|\sum_{c=1}^C\left(B_c^T \otimes I_n\right)\bar{x}_c\|^2\)) of \eqref{aug-lagrange}, can be decomposed, using the following lemma:

\begin{lemma}
\begin{equation}
\vectornorm{\sum_{c=1}^C\left(B_c^T \otimes I_n\right)\bar{x}_c}^2 = \sum_{j \in C_1}\left( D_j\vectornorm{x_j}_2^2 - \sum_{k \in N_j} x_j^Tx^k\right)
\end{equation}

and

\begin{equation}
\eta^T\sum_{c=1}^C\left(B_c^T \otimes I_n\right)\bar{x}_1 = \sum_{l\in C_c} \sum_{m\in N_l}sign\left(m-l\right)\eta_{ml}^T x_l
\end{equation}

where \(\eta\) is decomposed edge-wise: \(\eta = \left(\ldots, \eta_{ij},\ldots\right)\), such that \(\eta_{i,j} = \eta_{j,i}\), and is associated with the constraint \(x_i = x_j\).

\begin{proof}

\begin{align*}
u^Tu &= \sum	_{c_1=1}^C \sum	_{c_2=1}^C  \left(B_{c_1} \otimes x_{c_1}^T\right) \left(B_{c_2}^T \otimes x_c\right) \\
&= \sum_{c_1, c_2} B_{c_1}B_{c_2}^T \otimes x_{c_1}^Tx_{c_2}
\end{align*}

\(BB^T\) is a \(J \times J\) matrix, with the degree of the nodes on the main diagonal and \(-1\) in position \(\left(i,j\right)\) if nodes \(i\) and \(j\) are neighbours (i.e \(BB^T\) is the graph Laplacian). Hence, since we can write \(B_{c_1}B_{c_2}^T = w_{c_1}^TBB^Tw_{c_2}\), the trace of \(B_{c_1}B_{c_1}^T\) is simply the sum of the degrees of nodes with colour 1. 

For \(c_1 \neq c_2\),  \(B_{c_1}B_{c_2}^T\) corresponds to an off diagonal block of the graph Laplacian, and so counts how many neighbours each node with colour 1 has.

Finally, note that \(\eta \in \re^{nE}\) and can be written:

\begin{equation}
\eta = \sum_{c=1}^C w_c \otimes \eta_c
\end{equation}
where \(\eta_c\) is the vector of Lagrange multipliers associated across edges from colour \(c\). Now

\begin{align*}
\eta^Tu = \sum_{c_1=1}^C\sum_{c_2=1}^C w_{c_1}Bw_{c_2} \otimes \eta_{c_1}^Tx_c
\end{align*}
by the properties of Kronecker products, and the definition of \(B_c\). For \(c_1=c_2\), \(\eta^Tu\) is zero, as there are no edges between nodes of the same colour b definition. For \(c_1\neq c_2\), \(\eta^Tu\) counts the edges from \(c_1\) to \(c_2\), with the consideration that the edges from \(c_2\) to \(c_1\) are counted with opposite parity.
\end{proof}
\end{lemma}

Adding together this with the lemma, lets us write \eqref{aug-lagrange} as:

\begin{align}
L_\rho = \sum_{c=1}^C\sum_{j \in C_c} &\left(\|A_jx_j - y_j\|_2^2 + \beta\|z_j\|_1\right) + \nu^Tx_j \nonumber \\
& \text{        } + \frac{\rho}{2}D_i\vectornorm{x_j}^2 + \frac{\rho }{2}\|x_j-z_j\|^2
\label{generic-iterations}
\end{align}

where we have defined:

\begin{equation}
\nu_i = \left(\sum_{k \in \mathcal{N}_i} sign\left(k-i\right)\eta_{\{i,k\}} - \rho x_k \right)
\end{equation}

this is a rescaled version of the Lagrange multiplier, \(\eta\), which respects the graph structure. 

Then by differentiating \eqref{generic-iterations} with respect to \(x_j\) and \(z_j\) we  can find closed forms for the updates as:

\begin{thm}
\begin{align}
x_j^{k+1} &:= \left(A_j^TA_j + (\rho D_J + 1) I\right)^{-1}\left(A_j^Ty_j +  z^k - \nu^{kT}\right)\\
z_j^{k+1} &:= S_{\beta/\rho}\left(x_j^{k+1} \right)
 \\
\theta_j^{k+1} &:= \theta_j^{k} + \rho \left(x^{k+1}-z^{k+1}\right) \\
\eta_j^{k+1} &:= \eta_j^k + \rho\left(\sum_{m \in N_j} z_m^k - z_j^k\right)
\label{dadmm_algo_lasso}
\end{align}
\end{thm}

\section{Results} \label{sec:results}

\begin{figure}[h]
\centering
\includegraphics[height = 7.3 cm]{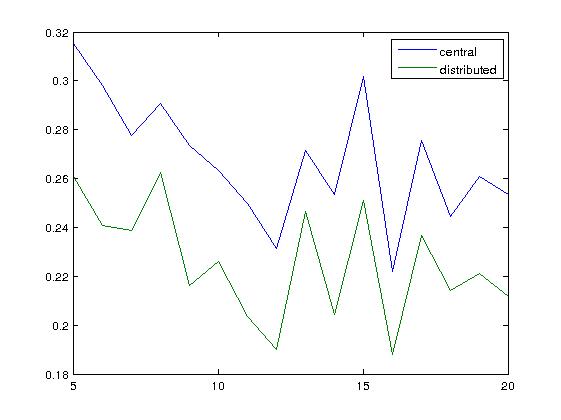}
\caption{Mse vs SNR for the sensing model, showing the performance of distributed and centralised solvers}
\label{msevssnr1}
\end{figure}

The model described in section (\ref{sec:sensingmodel}), equation \eqref{system} was simulated, with a wideband signal of 201 channels and a network of 50 nodes (i.e. the signal will be sampled at a 1/4 of rate predicted by Nyquist theory). The mixing patterns were generated from iid Gaussian sources (i.e the matrix S had each entry drawn from an iid Gaussian source). Monte Carlo simulations were performed at SNR values ranging from 5 to 20, and the expected Mean Squared Error (MSE) of solutions of a centralised solver (spgl1) and a distributed solver (ADMM) were calculated over 10 simulations per SNR value. The results can be seen in fig (\ref{msevssnr1}). 

The MSE was calculated as follows:

\begin{equation}
\frac{\vectornorm{Z^k - Z*}}{\vectornorm{Z*}}
\end{equation}

where \(Z^k\) is the result of the algorithm at iteration \(k\), and \(Z^*\) is the optimal solution.

These results indicate that for both centralised and distributed solvers, adding noise to the system results in a degrading of performance. Interestingly note, that the distributed solver seems to (slightly) outperform the centralised solver at all SNRs. This is counter-intuitive, as it would be expected that centralised solvers knowing \textit{all} the available information would outperform distributed solutions. We conjecture that the updates described in section \eqref{sec:opt-on-graphs}, take into account differences in noise across the network. The distributed averaging steps, which form the new prior for each node, then penalise updates from relatively more noisy observations. This corroborates observations from \cite{bazerque2008}.

This observation is (partially) confirmed in figure (\ref{erroriterations}), which plots the progress of the centralised and distributed solvers (as a function of iterations) towards the optimum solution. The SNR is 0.5 (i.e the signal is twice as strong as the noise). Note that after around 300 iterations, the MSE of the distributed solver is consistently below that of the centralised solver.

\begin{figure}[h]
\centering
\includegraphics[height = 7.3 cm]{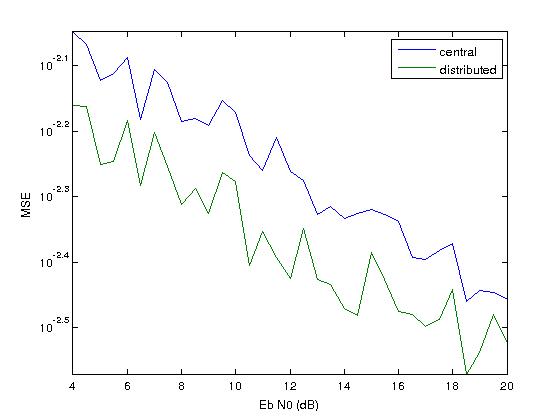}
\caption{Mse vs SNR for the sensing model, with AWGN only, showing the performance of distributed and centralised solvers}
\label{msevssnr0}
\end{figure}

\begin{figure}[h]
\centering
\includegraphics[height = 7.3 cm]{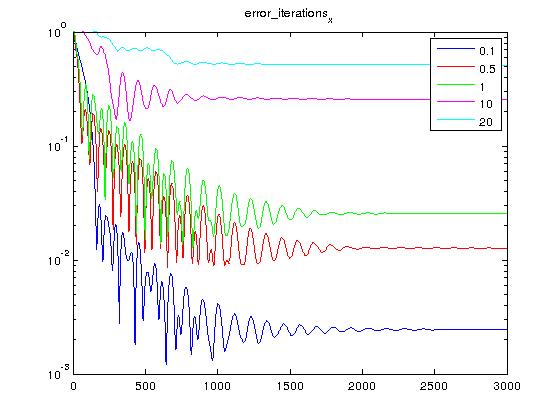}
\caption{The progress of the distributed solver as a function of the number of iterations, with different values of the regression parameter \(\lambda\)}
\label{fig:differentLambda}
\end{figure}

\begin{figure}[h]
\centering
\includegraphics[height = 7.3 cm]{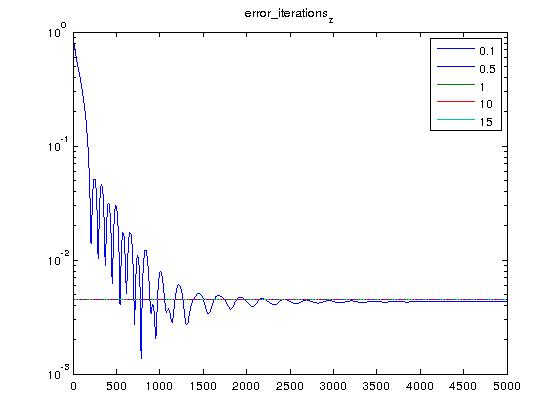}
\caption{The progress of a distributed (blue) and a centralised (green) solver as a function of the number of iterations. The value of \(\lambda = 0.1\)}
\label{fig:erroriterations}
\end{figure}

\section{Conclusions}
We have demonstrated an alternating direction algorithm for distributed optimisation with closed forms for the computation at each step, and discussed the statistical properties of the estimation. 

We have simulated the performance of this distributed algorithm for the distributed estimation of frequency spectra, in the presence of additive (white, Gaussian) and multiplicative (frequency flat) noise. We have shown that the algorithm is robust to a variety of SNRs and converges to the same solution as an equivalent centralised algorithm (in relative mean-squared-error).

We plan to work on larger, more detailed, models for the frequency spectra and to accelerate the convergence via Nesterov type methods to smooth the convergence of the distributed algorithm \cite{goldstein2014fast}. Specifically, we seek to dampen the ringing seen in Figure \ref{fig:erroriterations}

\bibliography{cswireless2}

\end{document}